\documentclass[10pt,twoside,BCOR7mm,DIV15,headinclude,footexclude,cleardoubleempty,idxtotoc]{scrartcl}

\usepackage[english]{babel}
\usepackage{graphicx}
\usepackage{hyperref}
\usepackage{scrpage2}
\usepackage{hyperref}
\usepackage{ifthen}

\makeatletter
\renewcommand{\@biblabel}[1]{}
\renewcommand{\@cite}[2]{%
{#1\ifthenelse{\boolean{@tempswa}}{,#2}{}}}
\makeatother

\hypersetup{breaklinks=true
,colorlinks=true,linkcolor=black,urlcolor=blue
,citecolor=black}

\ofoot{\thepage}
\ifoot{}

\setheadsepline{1pt}

\setkomafont{pagehead}{\normalfont\sffamily}
\setkomafont{pagenumber}{\normalfont\rmfamily}

\usepackage{booktabs}
\usepackage{amsmath}
\usepackage{amssymb}
\usepackage{multicol}
\usepackage{float}

\makeatletter
\newcommand{\listofcontributions}{\@starttoc{con}}

\newcommand{\l@contribution} {\@dottedtocline{1}{1.5em}{2.3em}}
\makeatother

\newenvironment{contribution}{
\setcounter{section}{0}
\setcounter{figure}{0}
\setcounter{table}{0}
\begin{flushleft}
{\em Clumping in Hot Star Winds \\
W.-R.\ Hamann, A.\ Feldmeier \& L.\ Oskinova, eds.\\
Potsdam: Univ.-Verl., 2007 \\
URN: http://nbn-resolving.de/urn:nbn:de:kobv:517-opus-13981
} 
\end{flushleft}
}{
\newpage
\lehead{}
\rohead{}
}

\begin{document}

\setlength{\baselineskip}{2.5ex}

\begin{contribution}

\lehead{Rich Townsend}

\rohead{Techniques for simulating radiative transfer through porous media}

\begin{center}
{\LARGE \bf Techniques for simulating radiative transfer through porous media}\\
\medskip

{\it\bf Rich Townsend}\\

{\it Bartol Research Institute, University of Delaware, Newark, DE 19716, USA}\\

\begin{abstract}
In this contribution, I discuss some basic techniques that can be used
to simulate radiative transfer through porous media. As specific
examples, I consider scattering transfer through a clumped slab, and
X-ray emission line formation in a clumped wind.
\end{abstract}
\end{center}

\begin{multicols}{2}

\section{Introduction}

A porous medium is one that is macroclumped: the typical scale of the
clumps is larger than the mean free path $\bar{\ell}$ of photons. This
represents the opposite end of the `clumping spectrum' to a
microclumped medium, for which the clump scale is smaller than
$\bar{\ell}$. The distinction between these two limits is more than
academic; although microclumping usually only affects the volume
emissivity of a medium (which depends on the local density squared),
porosity/macroclumping can result in the medium having an effective
opacity $\kappa_{\rm eff}$ that is significantly smaller than the
microscopic value $\kappa$ of the material comprising it.

Recent investigations of porosity (e.g., Oskinova, Feldmeier \& Hamann
\cite{townsend:Osk2004}; Owocki \& Cohen
\cite{townsend:OwoCoh2006}) have been focused primarily toward
evaluating its observable consequences. As such, these studies have
relied largely on very simple models for the radiative transfer. In
this contribution, I discuss some of the basic techniques and codes
that I have devised to move beyond these simplified models, allowing
consideration of the full radiative transfer problem for a clumped
medium.

The standard approach to radiative transfer is to solve the radiative
transfer equation (RTE),
\begin{equation} \label{townsend:rte}
\frac{{\rm d} I}{{\rm d} \tau} = I - S,
\end{equation}
where $I$ is the specific intensity, $S$ the source function, and
$\tau$ the optical depth coordinate. However, solving the RTE is not
the only way to tackle the problem. Instead of the $I$-based `field'
approach embodied by the RTE, one can adopt a `particle' approach,
following individual photons as they propagate through a medium. This
is precisely what Monte-Carlo radiative transfer entails, and in the
following section I apply this approach to model scattering transfer
through a clumped slab.

\section{A clumped scattering slab} \label{townsend:scattering}

To investigate radiative transfer in a clumped scattering medium, I
have developed \textsc{freyr}, a Monte-Carlo code for simulating the
propagation of photons through a plane-parallel slab composed of
discrete, spherical clumps. The clumps are randomly positioned, but
each has the same radius $r_{\rm c}$ and a uniform density $\rho_{\rm
c}$. Photons are introduced at the bottom of the slab with a random
upward direction. A nominal `optical distance to travel' $\tau$ is
assigned to each photon based on the expression
\begin{equation} \label{townsend:tau}
\tau = -\log x,
\end{equation}
where $x$ is a uniform random deviate in the range $0 < x \le 1$. This
optical distance is converted into a physical distance $\ell$ by
inverting the optical depth equation
\begin{equation} \label{townsend:optical-depth}
\tau = \int_{0}^{\ell} \kappa \rho(\ell')\, {\rm d} \ell';
\end{equation}
here, the microscopic opacity $\kappa$ is assumed to be constant,
while the local density $\rho$ varies with position as the photon
passes in and out of spheres. After moving a distance $\ell$ in the
appropriate direction, the photon is scattered: a new random direction
is chosen (corresponding to isotropic scattering), and a new optical
distance is picked using eqn.~(\ref{townsend:tau}). This procedure is
repeated until the photon escapes from the top or the bottom of the
slab; periodic boundary conditions are applied at the sides of the
slab.

Conceptually, this procedure is quite straightforward; but from an
implementation perspective the tricky part comes in inverting the
optical depth equation~(\ref{townsend:optical-depth}). The key to this
task lies in the recognition that along any given ray through the
slab, the density is a piecewise-constant function. Changes in $\rho$
arise only at those discrete points where the ray enters or exits a
sphere. To locate these points, the parametric equation for the ray
traversed by a photon is written as
\begin{equation}
{\bf r} = {\bf r}_{0} + \ell {\bf dr}.
\end{equation}

\begin{figure}[H]
\begin{center}
\includegraphics[width=\columnwidth]{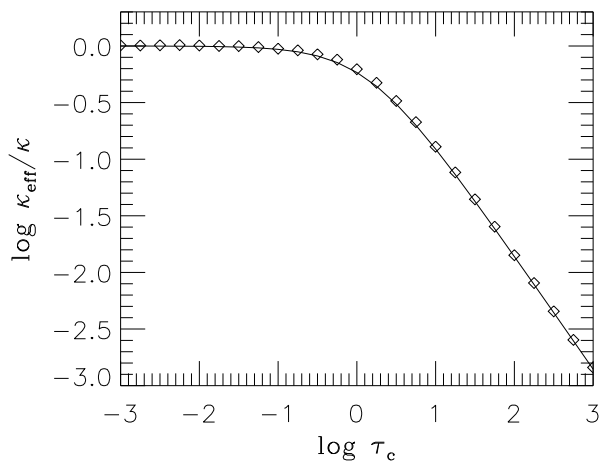}
\caption{The ratio between the effective opacity $\kappa_{\rm eff}$
and the microscopic opacity $\kappa$, plotted (diamonds) as a function of clump
optical thickness $\tau_{\rm c}$. The solid line shows the
corresponding prediction of the simple bridging
law~(\ref{townsend:bridge}).} \label{townsend:kappa-eff}
\end{center}
\end{figure}

Here, ${\bf r}_{0}$ is the starting position vector of the photon, and
the unit vector ${\bf dr}$ gives the photon's direction. To establish
whether this ray intersects a sphere located at ${\bf r}_{\rm s}$,
\textsc{freyr} calculates the following quantities:
\begin{align} \label{townsend:test-1}
{\bf v} &= {\bf r}_{0} - {\bf r}_{\rm s} \\
 \qquad B &= 2 {\bf dr} \cdot {\bf v} \\
 \qquad C &= {\bf v} \cdot {\bf v} - r_{\rm c}^{2} \\
  D &= B^2 - 4 C
\end{align}
If $D > 0$, then the ray pierces the sphere at the locations
\begin{equation} \label{townsend:test-2}
\ell_{\rm s} = \frac{-B \pm \sqrt{D}}{2};
\end{equation}
these correspond to the points where the density undergoes a
jump. Between these points, $\rho$ is constant, and $\tau$ varies
linearly -- so it is straightforward to reconstruct the function
$\tau(\ell)$ from eqn.~(\ref{townsend:optical-depth}), and then to
invert this function to find $\ell$ given $\tau$.

For each leg of a photon's journey, the ray-sphere intersection
test~(\ref{townsend:test-1}--\ref{townsend:test-2}) must be performed
against \emph{every} sphere contained in the slab. Potentially, this
can be quite time consuming, and so
\textsc{freyr} uses a spatial partitioning approach to reduce the
number of tests that must be done. After the distribution of spheres
is calculated (i.e., a set $\{{\bf r}_{\rm s}\}$ of random centers),
the slab is divided up into a 3-dimensional array of voxels. Each
voxel stores a index list of those spheres that fall part or all of
the way inside it. To find out which spheres a ray intersects, it is
then only necessary to test the subset of spheres that are indexed by
the voxels that the ray passes through. Efficient algorithms exist
(e.g., Amanatides \& Woo \cite{townsend:AmaWoo1987}) for quickly
finding these voxels.

Having described the basic principles of \textsc{freyr}, let me now
present some illustrative results. Fig.~\ref{townsend:kappa-eff} plots
the ratio $\kappa_{\rm eff}/\kappa$ between the effective and
microscopic opacities of a clumped slab, as a function of the averaged
clump optical thickness
\begin{equation}
\tau_{\rm c} = \frac{4 \kappa \rho_{\rm c} r_{\rm c}}{3}.
\end{equation}
To obtain the $\kappa_{\rm eff}$ datum at each $\tau_{\rm c}$
abscissa, a \textsc{freyr} simulation is used to determine an
empirical value for the slab transmittance
\begin{equation}
T \equiv \frac{N_{\rm esc}}{N_{\rm inj}};
\end{equation}
here, $N_{\rm inj}$ is the number of photons injected at the bottom of
the slab during the simulation, and $N_{\rm esc}$ is the number that
eventually escape through the top. Formally, the transmittance is a
function only of the slab optical thickness, so that
\begin{equation}
T = f(\kappa_{\rm eff} \bar{\rho} \Delta z)
\end{equation}
for some function $f()$. Thus, armed with the empirically-measured
value of $T$, and the known values of the slab mean density
$\bar{\rho}$ and vertical extent $\Delta z$, the effective opacity
follows as
\begin{equation}
\kappa_{\rm eff} = \frac{f^{-1}(T)}{\bar{\rho} \Delta z}.
\end{equation}
The only difficulty is to determine the inverse of the transmittance
function, $f^{-1}()$, but this can be done numerically using
Monte-Carlo simulations of a homogeneous, isotropically scattering
slab.

Looking at Fig.~\ref{townsend:kappa-eff}, the onset of porosity is
plain to see. Once the clumps become optically thick (i.e., $\tau_{\rm
c} > 1$), the effective opacity of the slab begins to fall well below
the microscopic value $\kappa$. This drop-off is well approximated by
\begin{equation} \label{townsend:bridge}
\frac{\kappa_{\rm eff}}{\kappa} = \frac{C}{C + \tau_{\rm c}}
\end{equation}
(shown in the figure by the solid line), where the constant $C
\approx 1.4$. This expression is a modified form of the bridging law
introduced by Cohen \& Owocki (\cite{townsend:OwoCoh2006}). In the
limit of small $\tau_{\rm c}$, it correctly reproduces the
microclumping case, for which $\kappa_{\rm eff}$ equals the
microscopic opacity $\kappa$. In the opposite limit of large
$\tau_{\rm c}$ (i.e., macroclumping), the effective opacity has a
scaling
\begin{equation}
\kappa_{\rm eff} = \frac{\kappa C}{\tau_{\rm c}} = \frac{3 C}{4 \rho_{\rm
c} r_{\rm c}}.
\end{equation}
Multiplying through by the clump mass $4\pi \rho_{\rm c} r_{\rm
c}^{3}/3$ then gives the effective cross section of the clumps,

\begin{figure}[H]
\begin{center}
\includegraphics[width=\columnwidth]{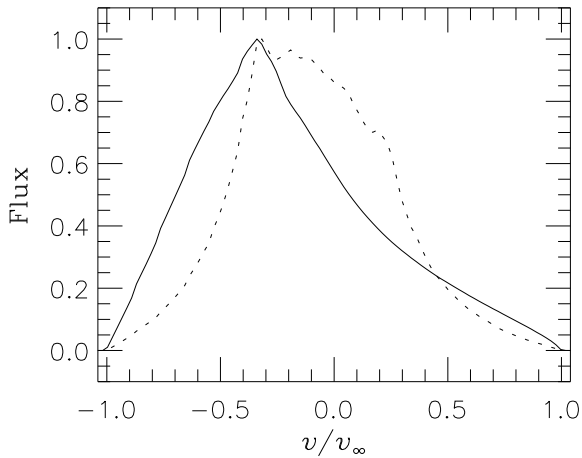}
\caption{X-ray line profiles for a $\tau_{\ast} = 3$ clumped wind. The
solid (dotted) profile corresponds to optically thin (optically thick)
clumps; both profiles are normalized to have a maximum flux of unity.}
\label{townsend:x-ray}
\end{center}
\end{figure}

\begin{equation}
\sigma_{\rm eff} = C \pi r_{\rm c}^{2}.
\end{equation}
This can be recognized as the geometric cross section $\pi r_{\rm
c}^{2}$, scaled by the constant $C$. The basis for this result is that
the clumps -- when they become extremely optically thick -- behave
like individual scattering centers, with an interaction cross section
equal to their geometric cross section. The appearance of the constant
$C$ is to correct for the angular scattering profile of each
individual clump. If this profile were isotropic, then we would obtain
$C=1$; but in fact the profile is that of a sphere with a Lambert-law
surface (see Schoenberg
\cite{townsend:Sch1929}), leading to the `observed' value $C = 13/9
\approx 1.4$.

\section{A clumped absorbing wind}

The utility of Monte-Carlo is that it works in cases such as
scattering where the source function $S$ is difficult to obtain. If
in fact $S$ is already available, then a simple formal solution of the
RTE (cf. eqn.~\ref{townsend:rte}) is
\begin{equation} \label{townsend:formal}
I = \int S {\rm e}^{-\tau} {\rm d}\tau.
\end{equation}
Calculating the intensity $I$ using this expression is invariably
faster than performing an equivalent Monte-Carlo simulation. (This is
a specific instance of a general rule: \emph{Monte-Carlo is guaranteed
to be the slowest way to solve a radiative transfer problem}. Of
course, sometimes -- as with scattering -- it is the only
straightforward way.)

The calculation of X-ray emission line profiles in clumped winds (see,
e.g., the contribution by David Cohen) is a specific example of a case
where the formal solution~(\ref{townsend:formal}) can be used. X-ray
line photons are emitted according to some simple parametric function
of radius, meaning that the source function is easy to
calculate. These photons are then re-absorbed by an opacity associated
with a distribution of clumps. As with the Monte-Carlo simulations,
the tricky part in modeling this re-absorption involves evaluating the
optical depth $\tau$.

Fortunately, for spherical clumps the same set of techniques
(cf.~\S\ref{townsend:scattering}) can be applied. If the inter-clump
medium is a vacuum, then once again the density along any given ray is
a piecewise-continuous function, and the ray-sphere intersection
testing (cf. eqns.~\ref{townsend:test-1}--\ref{townsend:test-2}) can
be used to reconstruct $\tau(\ell)$. If, more realistically, there is
a smooth inter-clump medium following a wind expansion law $\rho =
\dot{M}/(4 \pi r^{2} v)$, then the contribution of this medium to the
optical depth along a ray can be added in using numerical quadrature.

Fig.~\ref{townsend:x-ray} shows some example X-ray line profiles
calculated using \textsc{boreas}, a simple formal solver for a
spherical-clump wind. For both profiles, the optical depth
$\tau_{\ast}$ to the star (cf. Owocki \& Cohen
\cite{townsend:OwoCoh2006}) is 3; however, the profiles differ in the nature
of the clumping. The solid-line profile is for a configuration
composed of many optically thin clumps, and it shows the
characteristic asymmetric (blue-skewed) shape usually associated with
a smooth wind. Conversely, the dotted-line profile is for a
configuration composed of fewer, optically thick clumps. With porosity
acting to reduce the effective opacity of the wind, a much more
symmetric line profile is seen in this latter case.

\section{Summary}

The spherical clumps considered herein admittedly rather
idealized. However, it is precisely for such simple, idealized cases
that one can hope to fully understand the precise manner in which
porosity modifies the effective opacity of a medium. Both
\textsc{freyr} and
\textsc{boreas} -- and the techniques they implement -- will certainly
prove useful in developing such an understanding.


\end{multicols}

\end{contribution}


\end{document}